\documentclass[useAMS,usenatbib,onecolumn,usegraphicx]{mn2e}
\usepackage{pdfpages}

\usepackage{color}
\newcommand{\gsim}{\, \raisebox{-0.8ex}{$\stackrel{\textstyle >}{\sim}$ }}

\usepackage{ulem}

\newcommand{\beq}{\begin{equation}}
\newcommand{\eeq}{\end{equation}}
\newcommand{\beqar}{\begin{eqnarray}}
\newcommand{\eeqar}{\end{eqnarray}}

\title[Finite size effect in $\mu$]  % 33 letters
{Effects of finite sizes of atomic nuclei on shear modulus and torsional oscillations in neutron stars}
\author[H. Sotani, H. Togashi \& M. Takano]
{Hajime Sotani$^{1,2}$ \thanks{E-mail:sotani@yukawa.kyoto-u.ac.jp},
Hajime Togashi$^{3,4}$
and Masatoshi Takano$^{5,6}$
\\
$^1$Astrophysical Big Bang Laboratory, RIKEN, Saitama 351-0198, Japan\\
$^2$Interdisciplinary Theoretical \& Mathematical Science Program (iTHEMS), RIKEN, Saitama 351-0198, Japan\\
$^3$Department of Physics, Tohoku University, Sendai 980-8578, Japan\\
$^4$Waseda Research Institute for Science and Engineering, Waseda University, Tokyo 169-8555, Japan\\
$^5$Department of Physics, Waseda University, Tokyo 169-8555, Japan\\
$^6$Institute for Advanced Theoretical and Experimental Physics, Waseda University, Tokyo 169-8555, Japan}

\begin{document}
\maketitle
\label{firstpage}

%%%%%%%%%%%%%%%%%%%%%%%%%%%%%%%%%%%%%%%%%%%%%%%%
% Abstract
\begin{abstract}
The shear modulus of neutron star matter is one of the important properties for determining torsional oscillations in neutron stars. 
We take into account the effects of finite sizes of spherical nuclei on the shear modulus and examine the frequencies of crustal torsional oscillations. 
The shear modulus decreases owing to the finite-size effect, which in turn decreases the frequencies of torsional oscillations. 
In particular, the finite-size effect becomes more crucial for oscillations with a larger azimuthal quantum number and for neutron star models with a weaker density dependence of nuclear symmetry energy. 
In practice, when one identifies the quasi-periodic oscillations from a neutron star, 
where the magnetic effect is negligible, with crustal torsional oscillations, the finite-size effect can be more significant at frequencies higher than $\sim 100$ Hz. 
\end{abstract}
%%%%%%%%%%%%%%%%%%%%%%%%%%%%%%%%%%%%%%%%%%%%%%%%


%%%%%%%%%%%%%%%%%%%%%%%%%%%%%%%%%%%%%%%%%%%%%%%%
\section{Introduction}
\label{sec:I}
%%%%%%%%%%%%%%%%%%%%%%%%%%%%%%%%%%%%%%%%%%%%%%%%

A neutron star produced through a supernova explosion is one of the most suitable environments for probing the physics under extreme states. The density inside neutron stars easily exceeds the standard nuclear density, and the gravitational and magnetic fields can be much stronger than those observed in our solar system \citep{ST}. By observing neutron stars or their associated phenomena, one may obtain information that cannot be produced on Earth. 
For example, the discovery of 2$M_\odot$ neutron stars excludes all the soft equations of state (EOSs), with which the expected maximum mass does not reach the observed masses \citep{D10,A13,C20}. 
In addition, the pulsar light curve is affected by the general-relativistic light-bending effect, which primarily depends on stellar compactness, i.e., the ratio of the stellar mass to the stellar radius. 
Thus, by carefully observing the pulsar light curves, the mass and radius of neutron stars can be restricted \citep{Riley19,Riley21,Miller19,Miller21}. 
The resultant constraints are still largely uncertain, but they must be helpful in constraining the EOSs by accumulating the observational data on the mass and radius for additional objects. 
The gravitational wave event arising from the merging of binary neutron stars, GW170817 \citep{GW6,EM}, also provides information on the constraint on tidal deformability, which indicates that the $1.4M_\odot$ neutron star radius should be less than 13.6 km \citep{Annala18}.

The oscillation frequencies of neutron stars are another important piece of information. 
Since the oscillation spectrum strongly depends on the interior properties of an object, one may inversely obtain the interior information by observing the corresponding spectrum. 
This technique is known as asteroseismology. The asteroseismology of neutron stars was developed much earlier in order to interpret various possible neutron star oscillations 
detected in the electromagnetic spectrum (e.g., \cite{MVH88}). 
Even now, the most important results of asteroseismology are those associated with electromagnetic observations, e.g., quasi-periodic oscillations (QPOs) from soft-gamma repeaters (SGRs) as we will describe below. 
Furthermore, it has been proposed that the neutron star mass, radius, and EOS can be constrained by observing the gravitational waves from isolated compact objects (e.g., \cite{AK1996,AK1998,STM2001,SKH2004,SYMT2011,DGKK2013,Sotani21,SB21,SD22}).
Regarding the gravitational asteroseismology of neutron stars, the important role of $r$-modes in rapidly rotating neutron stars has also been often discussed (e.g., \cite{AK2001}). 
Additionally, to understand the gravitational wave signals appearing in the numerical simulation of a core-collapse supernova explosion, 
asteroseismology has recently been adopted (e.g., \cite{MRBV2018,SKTK2019,SS2019,TCPOF19,STT2021}). %Once the gravitational waves can be directly observed from neutron stars, one may be able to extract neutron star properties by asteroseismology, although they have never been detected yet. 
Gravitational waves from isolated neutron stars have never been detected yet, but once they can be directly observed, one may be able to extract neutron star properties by asteroseismology.
Instead of gravitational waves, however, several QPO frequencies have been detected in the x-ray afterglow following giant flares from SGRs, e.g., SGR 1900+14 and SGR 1806-20, whose frequencies are in the range of a few tenths of Hz up to kHz \citep{SW2005,SW2006}. If one assumes that these QPOs come from crustal torsional oscillations in a neutron star, one can constrain the crust properties (e.g., \cite{SW2009,GNHL2011,PA2012,SNIO2012,SNIO2013a,SNIO2013b,Sotani2014,SIO2016,SIO2017a,SIO2018,SIO2019}), even though the emission mechanism of QPOs is not understood yet. 
We note that, since the SGRs are magnetars, i.e., highly magnetized neutron stars, one may have to take into account the magnetic field effect on torsional oscillations. In fact, the frequencies of crustal torsional oscillations increase as the field strength increases \citep{Sotani2007,Gabler13}. Then, eventually, the crustal torsional oscillations disappear and Alfven oscillations become dominant as the magnetar oscillations. Nevertheless, considering the observational estimation of the magnetic field strength of SGR 1900+14 and SGR 1806-20 inferred from the spin-down rate \citep{K1998,H1999}, the crustal torsional oscillations must still survive.

Crustal torsional oscillations are characterized by the shear modulus in the crust region. The shear modulus is a property associated with rigidity against distortion. Therefore, it is mainly determined by the lattice energy in the crust. In the phase composed of spherical nuclei in the crust, where the nuclei form a body-centered cubic (bcc) lattice, the shear modulus for zero-temperature matter has been formulated as a function of the ion number density and charge of a nucleus \citep{SHOII1991} (see section \ref{sec:II} for details). 
However, in deriving the formula of the shear modulus, it is assumed that the nuclei in the crust are point-like particles. 
Indeed, this assumption may be valid for the nuclei in the outer crust, where neutrons are still confined in the nuclei, whereas the finite-size effect of nuclei must become non-negligible in the inner crust. Thus, in this paper, we first consider the shear modulus in the bcc lattice by including the finite-size effect of the nuclei and then discuss this effect on the frequency of crustal torsional oscillations in a nonmagnetized neutron star. The so-called pasta phase composed of nonspherical nuclei also contributes to crustal torsional oscillations (e.g., \cite{SIO2018,SIO2019}), but the thickness of such a phase is much smaller than that of the phase composed of spherical nuclei. Therefore, in this paper, we simply consider crustal torsional oscillations, whose restoring force comes from only the elasticity of the lattice structure with spherical nuclei. 
We note that the torsional oscillations are confined inside the crust region because the shear modulus vanishes inside the core region, where the matter behaves as a fluid.

%%%%%%%%%%%%%%%%%%%%%%%%%%%%%%%%%%%%%%%%%%%%%%%%
\section{Shear modulus}
\label{sec:II}
%%%%%%%%%%%%%%%%%%%%%%%%%%%%%%%%%%%%%%%%%%%%%%%%

The elasticity of the neutron star crust is characterized by the shear modulus. 
The shear modulus in the bcc lattice composed of spherical nuclei, which are the main components of the crust region, is approximately described in the limit of zero temperature as
\begin{equation}
  \mu_0 = 0.1194\,\frac{n_i (Ze)^2}{a},     \label{eq:mu0}
\end{equation}
where $n_i$, $Z$, and $a$ denote the ion number density, charge number of the ion, and the radius of a Wigner--Seitz cell, respectively, i.e., $4\pi a^3/3 = 1/n_i$ \citep{SHOII1991}. We note that this expression has been derived by averaging the overall direction of infinitesimal displacements with respect to the shear motion under the assumption that each spherical nucleus is only a point charge \citep{OI1990}. In Eq.~(\ref{eq:mu0}), one can observe that the shear modulus in the bcc lattice is proportional to the lattice (or Coulomb) energy. In fact, the shear modulus in two-dimensional geometry is directly expressed by the Coulomb energy in equilibrium \citep{PP1998}.

\begin{figure}
\begin{center}
\includegraphics[scale=0.5]{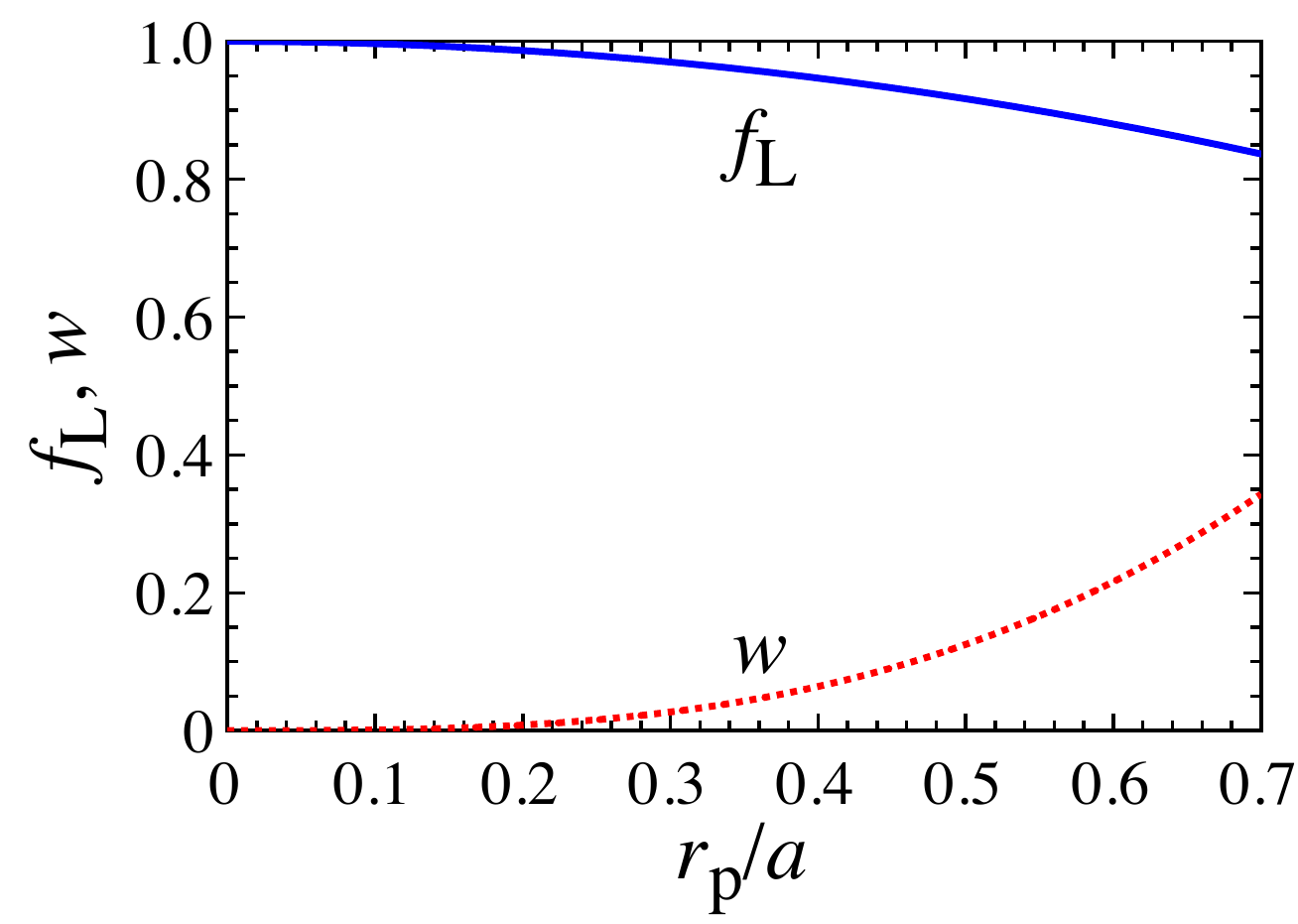} 
\end{center}
\caption{The contribution function of the finite-size effect of spherical nuclei ($f_{\rm L}$) and the volume fraction ($w$) are plotted as functions of the ratio of the reference proton radius to the radius of a Wigner--Seitz cell, where the solid and dotted lines correspond to $f_{\rm L}$ and $w$, respectively.}
\label{fig:fL}
\end{figure}

To take into account the size effect of spherical nuclei on the shear modulus, we assume the situation that the protons distribute uniformly with the density $n_{\rm p}$ in the region of $r\le r_{\rm p}$, where $r_{\rm p}$ is the reference proton radius defined by $4\pi r_{\rm p}^3n_{\rm p}/3=Z$. We note that the neutron radius $r_{\rm n}$ is generally not equivalent to $r_{\rm p}$, and the difference $r_{\rm n}-r_{\rm p}$ is the neutron skin thickness. 
Then, the lattice energy $W_{\rm L}$ of the unit cell within the Wigner--Seitz approximation for the compressible liquid drop model is written as \citep{BBP}
\begin{equation}
  W_{\rm L} = -\frac{9Z^2e^2}{10a}f_{\rm L},   \label{eq:lattice}
\end{equation}
where $f_L$ is a function of $r_{\rm p}/a$ given as
\begin{equation}
  f_{\rm L} = 1- \frac{1}{3}\left(\frac{r_{\rm p}}{a}\right)^2.  \label{eq:fl}
\end{equation}
Here, $W_{\rm L}$ in the limit of $r_{\rm p}=0$, where $f_L=1$, becomes the lattice energy with point-like nuclei, i.e., $f_L$ corresponds to a contribution function of the finite-size effect of spherical nuclei. 
In Fig. \ref{fig:fL}, we show the behavior of $f_{\rm L}$ and the volume fraction $w\equiv (r_{\rm p}/a)^3$ as a function of $r_{\rm p}/a$. From this figure, one can observe that the lattice energy slightly depends on the size of spherical nuclei. Since the ratio of $r_{\rm p}$ to $a$ (or the volume fraction) generally  increases with density, one can observe the stronger finite-size effect of the nuclei in the higher-density region. 
Since the shear modulus is directly associated with lattice energy as mentioned above, it would be natural to express the shear modulus including the finite-size effect of spherical nuclei, $\mu_f$, as
\begin{equation}
  \mu_f = \mu_0 f_{\rm L}. \label{eq:mu}
\end{equation}
Furthermore, nuclei may deform during the shear motion. 
If so, one has to take into account the effect of nuclear deformation on the shear modulus in the case of the finite-size nuclei. 
In fact, since the timescale of strong interactions (associated with the deformation) is much smaller than that of the Coulomb interaction, nuclei can deform, depending on their position. 
However, it is unclear how large the deformation of nuclei is during the shear motion because the amplitude of shear oscillations is not constrained by the observations. 
Thus, in this study, we simply adopt Eq. (\ref{eq:mu}) as the shear modulus in the case of finite-size nuclei.
We note that the phonon contribution to the shear modulus \citep{Baiko2011}, the electron screening effect on the shear modulus \citep{KP2013}, and the elastic properties of polycrystals \citep{KP2015} have been studied as other effects, but we simply neglect such effects in this study. 
We also note that the total Coulomb energy of a unit cell within the Wigner--Seitz approximation for a compressible liquid drop model is written as
\begin{equation}
  E_{\rm Coul} = \frac{3Z^2e^2}{5r_{\rm p}} + W_{\rm L} = \frac{3Z^2e^2}{5r_{\rm p}}f_{\rm C},   \label{eq:Coulomb}
\end{equation}
where $f_{\rm C}$ is a contribution function of the finite-size effect of spherical nuclei given as
\begin{equation}
  f_{\rm C} = 1- \frac{3}{2}\frac{r_{\rm p}}{a} + \frac{1}{2}\left(\frac{r_{\rm p}}{a}\right)^3.
\end{equation}
One may think that the shear modulus including the finite-size effect is constructed with $f_{\rm C}$ instead of $f_{\rm L}$ shown in Eq.~(\ref{eq:mu}), where
the dependence of $f_{\rm C}$ on $r_{\rm p}/a$ is stronger than that of $f_{\rm L}$. 
In this study, however, we employ $f_{\rm L}$ as shown in Eq.~(\ref{eq:mu}).

%%%%%%%%%%%%%%%%%%%%%%%%%%%%%%%%%%%
% Figure 1
%%%%%%%%%%%%%%%%%%%%%%%%%%%%%%%%%%%
%%%%%%%%%%%%%%%%%%%%%%%%%%%%%%%%%%%

\begin{figure}
\begin{center}
\includegraphics[scale=0.5]{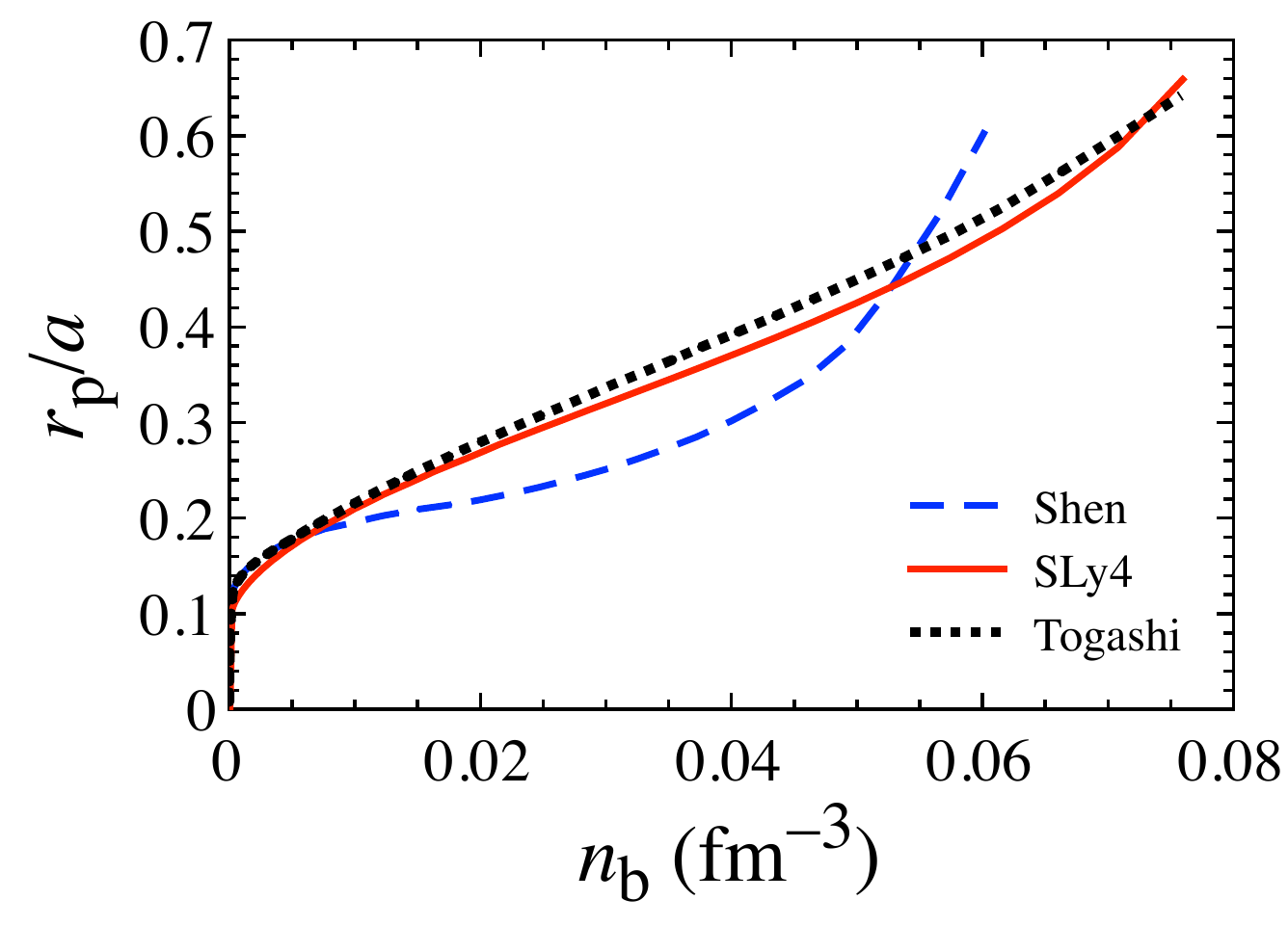} 
\end{center}
\caption{%%
The ratio $r_{\rm p}/a$ is shown as a function of the baryon number density for different EOSs, where the dashed, solid, and dotted lines correspond to the results obtained using the Shen, SLy4, and Togashi EOSs, respectively.
}%%
\label{fig:rpa}
\end{figure}

To examine how the frequencies of torsional oscillations depend on the EOSs, in this study, we adopt three different EOSs listed in Table~\ref{tab:EOS}, i.e., the EOS based on the relativistic mean-field (Shen EOS) \citep{Shen1998, Shen2011}, the EOS with the Skyrme-type interaction (SLy4 EOS) \citep{DH2001}, and the EOS derived by the variational method (Togashi EOS) \citep{Togashi17}. 
The derivation of the reference proton radius $r_{\rm p}$ used in Eq. (\ref{eq:fl}) for the SLy4 EOS is different from that for the other EOSs. 
That is, the SLy4 EOS was constructed within the compressible liquid drop model for non-uniform matter in the neutron star crust, 
so $r_{\rm p}$ can be strictly defined in the model. 
On the other hand, the Thomas--Fermi approximation was adopted to construct the Shen and Togashi EOSs for non-uniform matter, where the diffuseness of the nuclear surface was taken into account. 
For the Shen and Togashi EOSs, therefore, $r_{\rm p}$ used in this study is evaluated so that the root-mean-square proton radius calculated in the Thomas--Fermi approximation reproduces that in the compressible liquid drop model with $r_{\rm p}$. 
In Fig.~\ref{fig:rpa}, the ratio of $r_{\rm p}$ to $a$ for different EOSs is shown as a function of the baryon number density, where the dashed, solid, and dotted lines denote the results obtained using the Shen, SLy4, and Togashi EOSs, respectively. The values of $r_{\rm p}$ obtained using the Shen and Togashi EOSs are reasonable as compared with that obtained using the SLy4 EOS.

We note that the constraint on the $1.4M_\odot$ neutron star radius obtained from GW170817 has excluded the Shen EOS \citep{Annala18}, but we still consider this EOS to examine the EOS dependence of the frequencies of torsional oscillations. In Table~\ref{tab:EOS}, we list the EOS parameters, where $K_0$ is the incompressibility of the symmetric nuclear matter, $L$ is the density dependence of nuclear symmetry energy, and $\eta$ is given by $\eta\equiv (K_0L^2)^{1/3}$ (a suitable combination for expressing low-mass neutron stars proposed by \cite{SIOO14}). In addition, we also list the maximum mass and the transition density from the phase composed of spherical nuclei to uniform matter, i.e., the density at the bottom of the inner crust. We note that fission-like instability occurs when the volume fraction $w$ reaches $\sim 1/8$ \citep{PR1995}, which leads to the phase transition from the spherical nuclei phase to the so-called pasta phase and eventually to uniform matter, as the density increases. Even so, since the thickness of the pasta phase is much smaller than that of the phase composed of spherical nuclei \citep{SIO2017b}, we simply neglect the existence of the pasta phase in this study. 
This assumption is expected to be valid at least for the EOSs adopted in this study for the following reasons: 
The crustal torsional oscillations are confined inside the phase of spherical and cylindrical nuclei, since the slab-like nuclei behave as fluid against linear perturbation \citep{PP1998}. 
Furthermore, the frequencies inside the phase of spherical nuclei are almost the same as those in the phase of spherical and cylindrical nuclei when one considers the EOS with $L\gsim 40$ MeV, which was confirmed by \cite{SIO2018}.

%%%%%%%%%%%%%%%%%%%%%%%%%%%%%%
%   TABLE 1
%%%%%%%%%%%%%%%%%%%%%%%%%%%%%%
\begin{table}
\caption{EOS parameters adopted in this study, the maximum gravitational mass $M_{\rm max}$ for a cold neutron star constructed with the corresponding EOSs, and the density at the bottom edge of the neutron star crust, $n_{sp}$.} 
\label{tab:EOS}
\begin {center}
\begin{tabular}{cccccc}
\hline\hline
EOS & $K_0$ (MeV) & $L$ (MeV) & $\eta$ (MeV) & $M_{\rm max}/M_\odot$ & $n_{sp}$ (fm$^{-3}$)   \\
\hline
Shen    & 281  & 111  & 151  & 2.17  &  0.0602 \\
SLy4     & 230 & 45.9 & 78.5 & 2.05  & 0.0760   \\
Togashi & 245 & 38.7 & 71.6 & 2.21  & 0.0758 \\
\hline \hline
\end{tabular}
\end {center}
\end{table}
%%%%%%%%%%%%%%%%%%%%%%%%%%%%%%

\begin{figure}
\begin{center}
\includegraphics[scale=0.5]{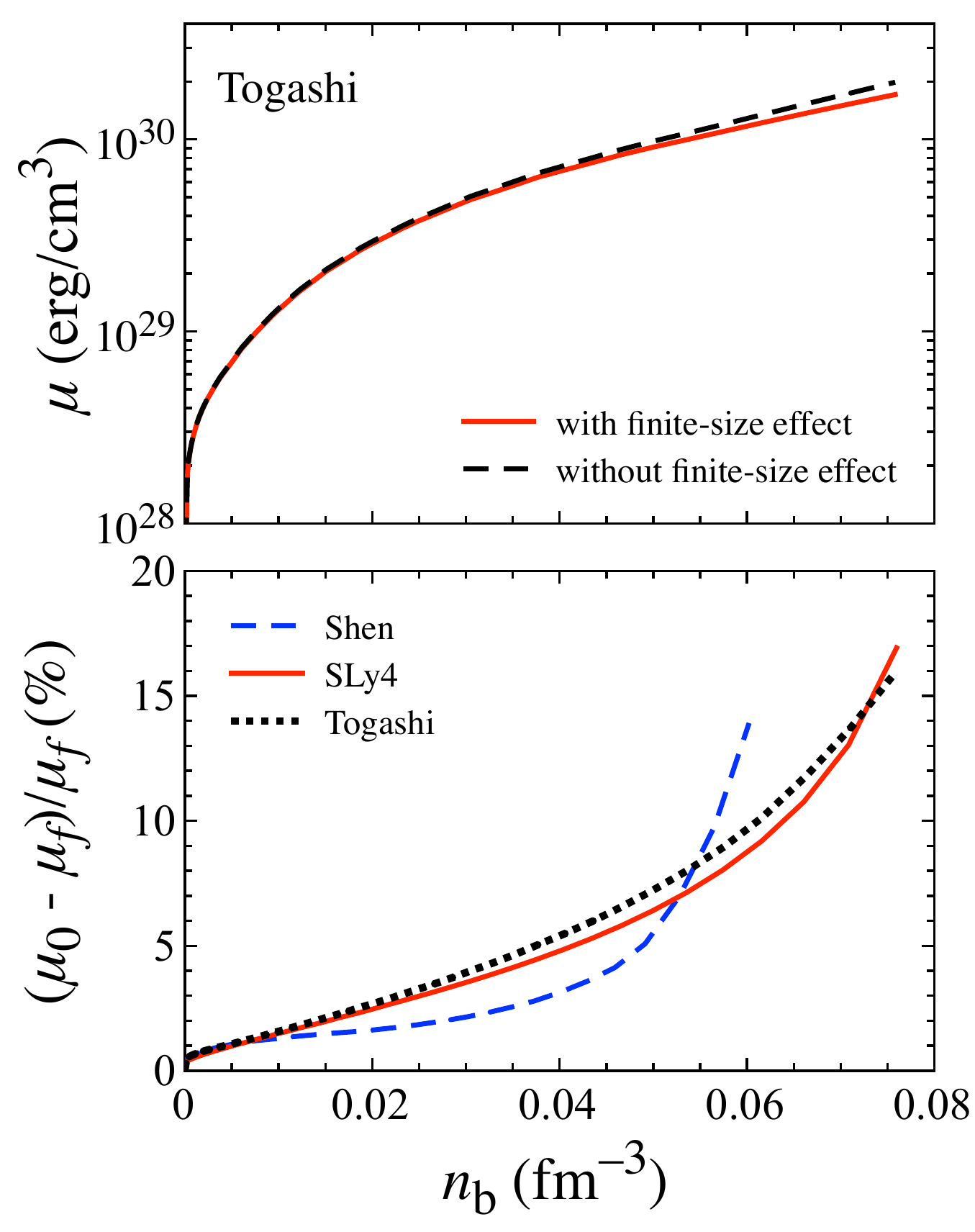} 
\end{center}
\caption{%%
Shear moduli with and without the finite-size effect are respectively shown with the solid and dashed lines as a function of the baryon number density for the Togashi EOS in the top panel. In the bottom panel, we show the relative deviation of the shear modulus without the finite-size effect ($\mu_0$) to that with the finite-size effect ($\mu_f$) for different EOSs, where the dashed, solid, and dotted lines correspond to the results obtained using the Shen, SLy4, and Togashi EOSs, respectively.
}%%
\label{fig:mu}
\end{figure}

In Fig.~\ref{fig:rpa}, the ratio of $r_{\rm p}$ to $a$ increases with density, which becomes more than $\sim 60\%$ at the bottom of the crust independently of the EOS. 
One also observes the EOS dependence (or $L$-dependence) in the profile of $r_{\rm p}/a$, i.e., $r_{\rm p}/a$ for an EOS with larger $L$ tends to be smaller.
This is because the symmetry energy at subnuclear density becomes smaller for an EOS with a larger $L$, which decreases the charge number of nuclei. 
The top panel in Fig. \ref{fig:mu} shows the shear modulus with and without the finite-size effect for the neutron star model constructed with the Togashi EOS;
$\mu_f$ becomes smaller than $\mu_0$ as the density increases owing to the finite-size effect.
The bottom panel in Fig. \ref{fig:mu} shows the relative deviations of the shear modulus for different EOSs; the deviations amount to $\sim 15\%$ at around the bottom of the crust.
We also note that the EOS dependence of the relative deviation shown in the bottom panel in Fig. \ref{fig:mu} is very similar to that of $r_{\rm p}/a$, as expected.

%%%%%%%%%%%%%%%%%%%%%%%%%%%%%%%%%%%
% Figure 2
%%%%%%%%%%%%%%%%%%%%%%%%%%%%%%%%%%%
%%%%%%%%%%%%%%%%%%%%%%%%%%%%%%%%%%%

%%%%%%%%%%%%%%%%%%%%%%%%%%%%%%%%%%%
% Figure 3
%%%%%%%%%%%%%%%%%%%%%%%%%%%%%%%%%%%
%%%%%%%%%%%%%%%%%%%%%%%%%%%%%%%%%%%

%%%%%%%%%%%%%%%%%%%%%%%%%%%%%%%%%%%%%%%%%%%%%%%%
\section{Torsional oscillations}
\label{sec:III}
%%%%%%%%%%%%%%%%%%%%%%%%%%%%%%%%%%%%%%%%%%%%%%%%

In this study, we simply consider a neutron star without a magnetic field.
As mentioned above, since the torsional oscillations excited in a nonmagnetized neutron star are confined inside the neutron star crust, we focus only on the crust region. 
Spherically symmetric neutron star models are generally constructed by integrating the Tolman--Oppenheimer--Volkoff (TOV) equations together with an appropriate EOS from the center up to the surface outward, for a given central density. 
On the other hand, to eliminate uncertainties in the EOS for the higher-density region, which corresponds to the neutron star core region, in this study, we construct the neutron star crust by integrating the TOV equations inward from the stellar surface down to the bottom edge of the crust for selected values of stellar mass and radius, as in \cite{SNIO2012,SNIO2013a}. 
In such a crust model, we consider the torsional oscillations.

Since the crustal torsional oscillations are of the axial type, the relativistic Cowling approximation works well, i.e., the metric perturbations during the oscillations can be omitted. The perturbation equation for the angular displacement can be derived from the linearized equation of motion \citep{ST83}. Since the core region of a neutron star behaves as a fluid, torsional oscillations are confined only to the crust region. Thus, the boundary conditions are imposed on the bottom edge of the crust and at the stellar surface. In practice, we adopt the zero-traction condition at the bottom edge of the crust and the zero-torque condition at the stellar surface \citep{ST83,Sotani2007}. The concrete perturbation equation and boundary conditions are shown in \cite{SNIO2012}. Then, the problem to solve becomes an eigenvalue problem.

\begin{figure}
\begin{center}
\includegraphics[scale=0.5]{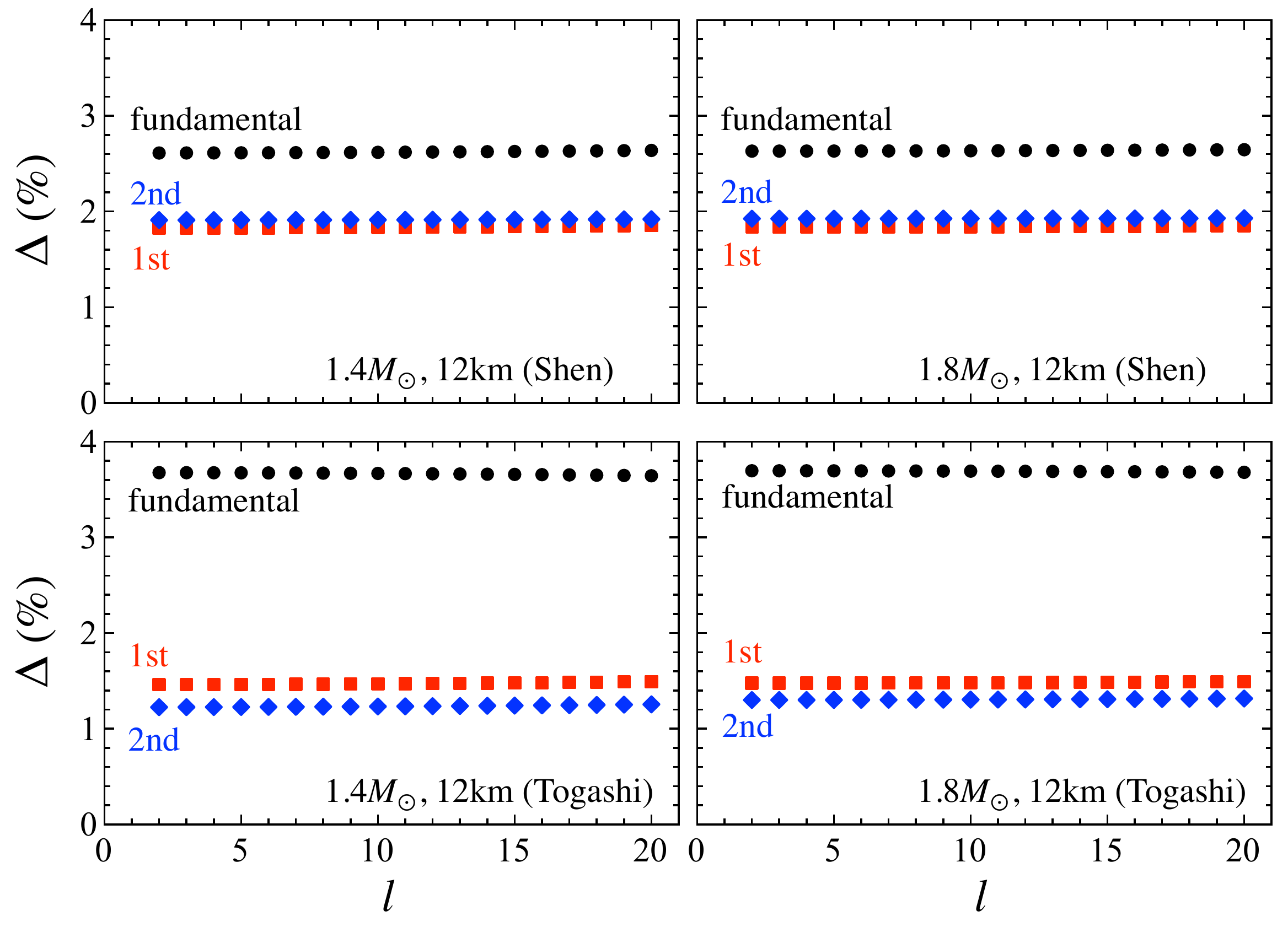} 
\end{center}
\caption{%%
The relative deviation of frequencies without the finite-size effect from those with such an effect, defined by Eq. (\ref{eq:delta}), is shown as a function of $\ell$, where the circles, squares, and diamonds denote the results for the fundamental oscillations ($n=0$) and the 1st ($n=1$) and 2nd overtones $(n=2)$, respectively. The left and right panels correspond to the neutron star model with $1.4M_\odot$ and $1.8M_\odot$, whereas the top and bottom panels correspond to the neutron star model constructed from the results obtained using the Shen and Togashi EOSs, respectively. 
}%%
\label{fig:dt}
\end{figure}

As shown in Fig. \ref{fig:mu}, the shear modulus decreases owing to the finite-size effect. 
As a result, since the shear velocity given by $(\mu/(\varepsilon+p))^{1/2}$ with the energy density $\varepsilon$ and pressure $p$ also decreases, one can expect that the frequencies of torsional oscillations with the finite-size effect become smaller than those without the effect. 
In Fig. \ref{fig:dt}, we show the relative deviation $\Delta$ given as 
\begin{equation}
  \Delta = \frac{{}_\ell f_n^{(-)} - {}_\ell f_n^{(+)}}{{}_\ell f_n^{(+)}}, \label{eq:delta}
\end{equation}
where ${}_\ell f_n^{(+)}$ and ${}_\ell f_n^{(-)}$ respectively correspond to the frequencies of the $\ell$-th torsional oscillations with and without the finite-size effect, while $n$ denotes the nodal number in the eigenfunction. 
Since the results obtained using the SLy4 EOS are almost the same as those obtained using the Togashi EOS, hereafter we focus only on the results obtained using the Shen and Togashi EOSs. From this figure, one can observe that the frequencies decrease owing to the finite-size effect, as expected, but the relative deviation $\Delta$ is almost independent of $\ell$. 
One also finds that the relative deviation in the fundamental frequencies $(n=0)$ decreases for the neutron star model with a larger $L$, which is caused by the $L$-dependence in the shear modulus as shown in Fig. \ref{fig:mu}, whereas the relative deviation in overtones does not seem to be very simple because the frequencies of overtones depend on not only the shear modulus but also the crust thickness \citep{HC80}\footnote{We also note that, for strongly magnetized neutron stars, the frequency modulation due to the fine-size effect of atomic nuclei as shown in Fig. \ref{fig:dt} is much smaller than that due to the magnetic fields (e.g., \cite{Sotani2007,Gabler13}).}. 
However, the fundamental oscillations may be more important, if one is going to identify the QPO frequencies from a neutron star with crustal torsional oscillations.
In fact, the frequencies observed in magnetar giant flares, such as SGR 1806-20 or SGR 1900$+$14, range from $\sim 10$ Hz up to the order of kilohertz, 
among which the frequencies of about 100 Hz, e.g., 150 Hz in SGR 1806-20 or 155 Hz in SGR 1900+14, are the most suitable ones to be identified with the fundamental frequencies of the crustal torsional oscillations for certain values of $\ell$
(see \cite{SNIO2012,SNIO2013a}). 
In the following, therefore, we focus on fundamental oscillations.

%%%%%%%%%%%%%%%%%%%%%%%%%%%%%%%%%%%
% Figure 4
%%%%%%%%%%%%%%%%%%%%%%%%%%%%%%%%%%%
%%%%%%%%%%%%%%%%%%%%%%%%%%%%%%%%%%%

\begin{figure}
\begin{center}
\includegraphics[scale=0.5]{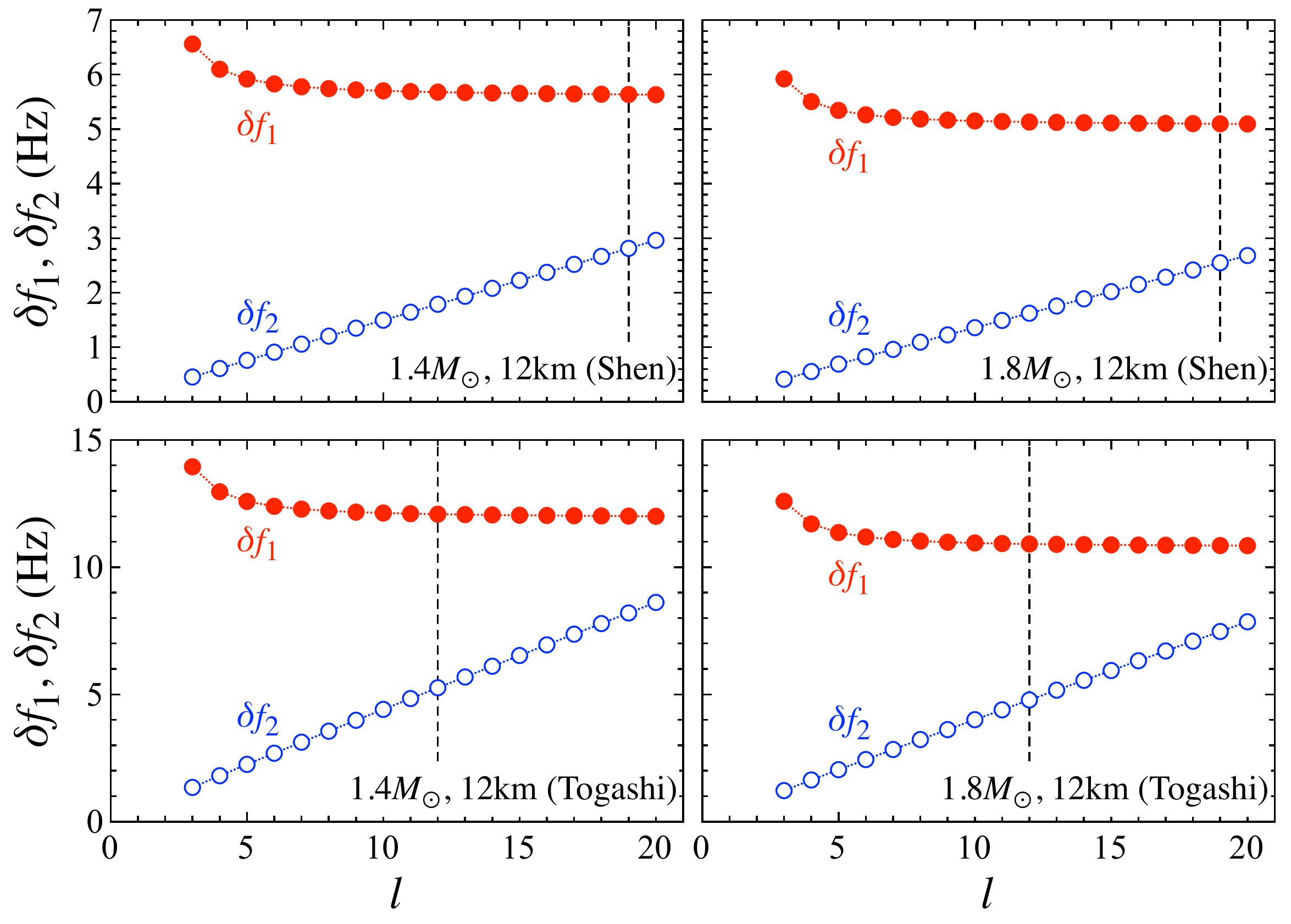} 
\end{center}
\caption{%%
The deviations of the frequencies, $\delta f_1$ and $\delta f_2$, are shown as a function of $\ell$, where $\delta f_1$ and $\delta f_2$ are given by Eqs. (\ref{eq:df1}) and (\ref{eq:df2}), respectively. The vertical dashed line denotes the values of $\ell$ at which the ratio of $\delta f_2$ to $\delta f_1$, i.e., $\delta f_2/\delta f_1$, becomes more than 0.5.
}%%
\label{fig:df}
\end{figure}

Although the relative deviation $\Delta$ is almost independent of $\ell$ as shown in Fig. \ref{fig:dt}, the absolute value of the deviation of frequency increases with $\ell$ because the fundamental frequency itself increases with $\ell$. 
To choose the value of $\ell$ above which the finite-size effect is not negligible, we examine the following two quantities, i.e., $\delta f_1$, which is the deviation between the $\ell$-th and $(\ell-1)$-th fundamental frequencies without the finite-size effect, and $\delta f_2$, which is the deviation between the $\ell$-th fundamental frequencies without and with the finite-size effect.
Explicitly, they are given by 
\begin{eqnarray}
  \delta f_1(\ell) &=& {}_\ell f_0^{(-)} - {}_{\ell-1}f_0^{(-)}, \label{eq:df1}  \\
  \delta f_2(\ell) &=& {}_\ell f_0^{(-)} - {}_{\ell}f_0^{(+)}.   \label{eq:df2}
\end{eqnarray}
The resultant values of $\delta f_1$ and $\delta f_2$ are shown in Fig. \ref{fig:df}, where the filled and open circles respectively correspond to $\delta f_1$ and $\delta f_2$. 
As seen in this figure, $\delta f_2$ increases with $\ell$, which means that the finite-size effect becomes more important as $\ell$ increases. 
In fact, we show the $\ell$ value above which $\delta f_2/\delta f_1$ becomes larger than 0.5 with the vertical dashed line in Fig. \ref{fig:df} for reference. 
For the neutron star model constructed using the Togashi, SLy4, and Shen EOSs, the corresponding $\ell$ values are 12, 13, and 19, respectively. 
This implies that the finite-size effect becomes more important for the stellar model with a smaller $L$.

%%%%%%%%%%%%%%%%%%%%%%%%%%%%%%%%%%%
% Figure 5
%%%%%%%%%%%%%%%%%%%%%%%%%%%%%%%%%%%
\section{Conclusion}
\label{sec:IV}
%%%%%%%%%%%%%%%%%%%%%%%%%%%%%%%%%%%%%%%%%%%%%%%%

The QPOs observed in the afterglow following magnetar giant flares are considered to be strongly associated with neutron star oscillations. 
One of the possible theoretical explanations of the observations is crustal torsional oscillations, in which the oscillation frequencies are governed by the shear modulus. 
The shear modulus in the bcc lattice composed of spherical nuclei is well known and formulated under the assumption that the nuclei are point-like particles. 
In this study, we focus on the finite-size effect of spherical nuclei and take it into account the value of the shear modulus. 
The shear modulus decreases owing to the finite-size effect, which is more important for EOSs with a smaller $L$. We also examine the frequencies of crustal torsional oscillations in a nonmagnetized neutron star to evaluate the finite-size effect. 
We find that the finite-size effect becomes more crucial in torsional oscillations with a larger $\ell$ and in the neutron star models constructed using the EOSs with a smaller $L$. In practice, if one identifies the observed QPO frequencies with crustal torsional oscillations, the finite-size effect is not negligible for QPOs of more than $\sim 100$ Hz.
We note that the QPO frequencies detected in the afterglow of magnetar giant flares can be more strongly affected by magnetic fields (e.g., \cite{Sotani2007,Gabler13}) than by the finite-size effects. Therefore, the latter effects would be more pronounced if QPOs associated with torsional oscillations were detected from weakly magnetized neutron stars.

%\newpage
%%%%%%%%%%%%%%%%%%%%%%%%%%%%%%%%%%%%%%%%%%%%%%%%
%\acknowledgments
%%%%%%%%%%%%%%%%%%%%%%%%%%%%%%%%%%%%%%%%%%%%%%%%
\section*{Acknowledgements}
We would like to express our special thanks to H. Shen for providing us with numerical data on parameters of the Thomas--Fermi calculation. 
%H.S. is grateful to S. Furusawa, A. Ohnishi, and some members in the Division of Theoretical Astronomy, National Astronomical Observatory of Japan for discussing the lattice energy in the crust region. 
This work was supported in part by Japan Society for the Promotion of Science (JSPS) KAKENHI Grant Numbers 
JP18K13551,  % Wakate by Togashi
JP19KK0354,  % International (A) by Sotani
JP20H04753,  % Koubo Genesis by Sotani
JP21H01088, and % Kiban(B) by Sotani
JP21K13924   % Wakate by Togashi
and by Pioneering Program of RIKEN for Evolution of Matter in the Universe (r-EMU).

\section*{Data availability}
The data underlying this article will be shared on reasonable request to the corresponding author.

%\appendix
%%%%%%%%%%%%%%%%%%%%%%%%%%%%%%%%%%%%%%%%%%%%%%%%
%\section{Stellar Models with Dipole Magnetic field}   % Appendix A
%\label{sec:appendix_1}
%%%%%%%%%%%%%%%%%%%%%%%%%%%%%%%%%%%%%%%%%%%%%%%%

%%%%%%%%%%%%%%%%%%%%%%%%%%%%%%%%%%%%%%%%%%%%%%%%


\begin{thebibliography}{999}
%%%%%%%%%%%%%%%%%%%%%%%%%%%%%%%%%%%%%%%%%%%%%%%%

\bibitem[\protect\citeauthoryear{Abbott et al.}{2017a}]{GW6}  % GW170817
   Abbott B. P. et al. (LIGO Scientific Collaboration and Virgo Collaboration), 2017a, Phys. Rev. Lett., 119, 161101

\bibitem[\protect\citeauthoryear{Abbott et al.}{2017b}]{EM}  % EM in GW170817
   Abbott B. P. et al. (LIGO Scientific Collaboration and Virgo Collaboration), 2017b, ApJ, 848, L12

\bibitem[\protect\citeauthoryear{Andersson \& Kokkotas}{1996}]{AK1996}
   Andersson N., Kokkotas K. D., 1996, Phys.\ Rev.\ Lett., 677, 4134

\bibitem[\protect\citeauthoryear{Andersson \& Kokkotas}{1998}]{AK1998}
   Andersson N., Kokkotas K. D., 1998, MNRAS, 299, 1059

\bibitem[\protect\citeauthoryear{Andersson \& Kokkotas}{2001}]{AK2001}
   Andersson N., Kokkotas K. D., 2001, Int. J. Mod. Phys. D, 10, 381

\bibitem[\protect\citeauthoryear{Annala et al.}{2018}]{Annala18}  % R1.4 <= 13.6km
   Annala E., Gorda T., Kurkela A., Vuorinen A., 2018, Phys. Rev. Lett., 120, 172703

\bibitem[\protect\citeauthoryear{Antoniadis et al.}{2013}]{A13}
   Antoniadis J. et al., 2013, Science, 340, 6131

\bibitem[\protect\citeauthoryear{Baiko}{2011}]{Baiko2011}
   Baiko D. A., 2011, MNRAS, 416, 22
   
\bibitem[\protect\citeauthoryear{Baym, Bethe \& Pethick}{1971}]{BBP}
   Baym G., Bethe H. A., Pethick C. J., 1971, Nucl. Phys. A, 175, 225

% M = 2.14(+0.20-0.18)\UTF{2009}M_sun (MSP J0740+6620)
\bibitem[\protect\citeauthoryear{Cromartie et al.}{2020}]{C20}    
   Cromartie H. T. et al., 2020, Nat. Astron., 4, 72

\bibitem[\protect\citeauthoryear{Demorest et al.}{2010}]{D10} 
   Demorest P., Pennucci T., Ransom S., Roberts M., Hessels J., 2010, Nature, 467, 1081

\bibitem[\protect\citeauthoryear{Doneva et al.}{2013}]{DGKK2013}
   Doneva D. D., Gaertig E., Kokkotas K. D., Kr\"{u}ger C., 2013, Phys.\ Rev.\ D, 88, 044052

\bibitem[\protect\citeauthoryear{Douchin \& Haensel}{2001}]{DH2001}
   Douchin F., Haensel P., 2001, A\&A, 380, 151

\bibitem[\protect\citeauthoryear{Gabler et al.}{2013}]{Gabler13}
   Gabler M., Cerd\'{a}-Dur\'{a}n P., Font J. A., M\"{u}ller E., Stergioulas N., 2013, MNRAS, 430, 1811

\bibitem[\protect\citeauthoryear{Gearheart et al.}{2011}]{GNHL2011}
   Gearheart M., Newton W. G., Hooker J., Li B. A., 2011, MNRAS, 418, 2343

%\bibitem[\protect\citeauthoryear{Haensel, Potekhin \& Yakovlev}{2006}]{NS}
%   Haensel P., Potekhin A. Y., Yakovlev D. G., in
%   {\it Neutron Stars 1: Equation of State and Structure} (Springer, 2006)

\bibitem[\protect\citeauthoryear{Hansen \& Cioffi}{1980}]{HC80}
   Hansen C., Cioffi D. F., 1980, ApJ, 238, 740

%\bibitem[\protect\citeauthoryear{Horowitz \& Hughto}{2008}]{HH2008}
%   Horowitz C. J., Hughto J., 2008, preprint (arXiv:0812.2650)

\bibitem[\protect\citeauthoryear{Hurley et al.}{1999}]{H1999}
   Hurley K. et al., 1999, ApJ, 510, L111

%\bibitem[\protect\citeauthoryear{Israel et al.}{2005}]{I2005} 
%   Israel G. et al., 2005 Astrophys. J., 628, L53 

% electron screening
\bibitem[\protect\citeauthoryear{Kobyakov \& Pethick}{2013}]{KP2013}
   Kobyakov D., Pethick C. J., 2013, Phys. Rev. C, 87, 055803

% Elastic properties of polycrystalline
\bibitem[\protect\citeauthoryear{Kobyakov \& Pethick}{2015}]{KP2015}
   Kobyakov D., Pethick C. J., 2015, MNRAS, 449, L110

\bibitem[\protect\citeauthoryear{Kouveliotou et al.}{1998}]{K1998}
   Kouveliotou C. et al., 1998, Nature, 393, 235

\bibitem[\protect\citeauthoryear{McDermott, Van Horn, \& Hansen}{1988}]{MVH88}
   McDermott P. N., Van Horn H. M., Hansen C. J., 1988, ApJ, 325 725

\bibitem[\protect\citeauthoryear{Miller et al.}{2019}]{Miller19} 
   Miller M. C. et al., 2019, ApJ, 887, L24

\bibitem[\protect\citeauthoryear{Miller et al.}{2021}]{Miller21} 
   Miller M. C. et al., 2021, ApJ, 918, L28

\bibitem[\protect\citeauthoryear{Morozova et al.}{2018}]{MRBV2018}
  Morozova V., Radice D., Burrows A., Vartanyan D., 2018, ApJ, 861, 10

\bibitem[\protect\citeauthoryear{Ogata \& Ichimaru}{1990}]{OI1990}
   Ogata S., Ichimaru S., 1990, Phys.\ Rev.\ A, 42, 4867

\bibitem[\protect\citeauthoryear{Passamonti \& Andersson}{2012}]{PA2012}
   Passamonti A., Andersson N., 2012, MNRAS, 419, 638
%  Towards real neutron star seismology: Accounting for elasticity and superfluidity (without magnetic fields)

\bibitem[\protect\citeauthoryear{Pethick \& Potekhin}{1998}]{PP1998}
   Pethick C. J., Potekhin A. Y., 1998, Phys. Lett. B, 427, 7

\bibitem[\protect\citeauthoryear{Pethick \& Ravenhall}{1995}]{PR1995}
   Pethick C. J., Ravenhall D. G., 1995, Annu. Rev. Nucl. Part. Sci., 45, 429

\bibitem[\protect\citeauthoryear{Riley et al.}{2019}]{Riley19} 
   Riley T. E. et al., 2019, ApJ, 887, L21

\bibitem[\protect\citeauthoryear{Riley et al.}{2021}]{Riley21} 
   Riley T. E. et al., 2021, ApJ, 918, L27

\bibitem[\protect\citeauthoryear{Schumaker \& Thorne}{1983}]{ST83} 
   Schumaker B. L., Thorne K. S., 1983, MNRAS, 203, 457

\bibitem[\protect\citeauthoryear{Shapiro \& Teukolsky}{1983}]{ST}
   Shapiro S. L., Teukolsky S. A., in
   {\it Black Holes, White Dwarfs, and Neutron Stars} (A Wiley-Interscience Publication, 1983)

\bibitem[\protect\citeauthoryear{Shen et al.}{1998}]{Shen1998}  % Shen
   Shen H., Toki H., Oyamatsu K., Sumiyoshi K., 1998, Nucl. Phys. A, 637, 435
   
\bibitem[\protect\citeauthoryear{Shen et al.}{2011}]{Shen2011}  % Shen
   Shen H., Toki H., Oyamatsu K., Sumiyoshi K., 2011, ApJS, 197, 20

%\bibitem[\protect\citeauthoryear{Sotani \& Harada}{2003}]{SH2003}
%   Sotani H., Harada T., 2003, Phys.\ Rev.\ D, 68, 024019

%\bibitem[\protect\citeauthoryear{Sotani, Kokkotas \& Stergioulas}{2008}]{Sotani2008}
%   Sotani H., Kokkotas K. D., Stergioulas N., 2008, MNRASL, 385, L5

%\bibitem[\protect\citeauthoryear{Sotani, Colaiuda \& Kokkotas}{2008}]{Sotani2008b}
%   Sotani H., Colaiuda A., Kokkotas K. D., 2008, MNRAS, 385, 2161

%\bibitem[\protect\citeauthoryear{Sotani \& Kokkotas}{2009}]{Sotani2009}
%   Sotani H., Kokkotas K. D., 2009, MNRAS, 395, 1163

%\bibitem[\protect\citeauthoryear{Sotani}{2011}]{Sotani2011}
%   Sotani H., 2011, MNRAS, 417, L70

\bibitem[\protect\citeauthoryear{Sotani}{2014}]{Sotani2014}
   Sotani H., 2014, Phys. Lett. B, 730, 166

%\bibitem[\protect\citeauthoryear{Sotani}{2020a}]{Sotani20a}
%   Sotani H., 2020, Phys. Rev. D, 102, 063023

%\bibitem[\protect\citeauthoryear{Sotani}{2020b}]{Sotani20b}
%   Sotani H., 2020, Phys. Rev. D, 102, 103021

\bibitem[\protect\citeauthoryear{Sotani}{2021}]{Sotani21}
   Sotani H., 2021, Phys. Rev. D, 103, 123015

\bibitem[\protect\citeauthoryear{Sotani \& Kumar}{2021}]{SB21}
   Sotani H., Kumar B., 2021, Phys. Rev. D, 104, 123002

\bibitem[\protect\citeauthoryear{Sotani \& Dohi}{2022}]{SD22}
   Sotani H., Dohi A., 2022, Phys. Rev. D, 105, 023007 

\bibitem[\protect\citeauthoryear{Sotani, Iida \& Oyamatsu}{2016}]{SIO2016}
   Sotani H., Iida K., Oyamatsu K., 2016, New Astron., 43, 80
 
\bibitem[\protect\citeauthoryear{Sotani, Iida \& Oyamatsu}{2017a}]{SIO2017a}
   Sotani H., Iida K., Oyamatsu K., 2017a, MNRAS, 464, 3101

\bibitem[\protect\citeauthoryear{Sotani, Iida \& Oyamatsu}{2017b}]{SIO2017b}
   Sotani H., Iida K., Oyamatsu K., 2017b, MNRAS, 470, 4397

\bibitem[\protect\citeauthoryear{Sotani, Iida \& Oyamatsu}{2018}]{SIO2018}
   Sotani H., Iida K., Oyamatsu K., 2018, MNRAS, 479, 4735
   
\bibitem[\protect\citeauthoryear{Sotani, Iida \& Oyamatsu}{2019}]{SIO2019}
   Sotani H., Iida K., Oyamatsu K., 2019, MNRAS, 489, 3022

\bibitem[\protect\citeauthoryear{Sotani et al.}{2014}]{SIOO14} 
   Sotani H., Iida K., Oyamatsu K., Ohnishi A., 2014, Prog. Theor. Exp. Phys. 2014, 051E01

\bibitem[\protect\citeauthoryear{Sotani, Kohri \& Harada}{2004}]{SKH2004}
   Sotani H., Kohri K., Harada T., 2004, Phys.\ Rev.\ D, 69, 084008

\bibitem[\protect\citeauthoryear{Sotani, Kokkotas \& Stergioulas}{2007}]{Sotani2007}
   Sotani H., Kokkotas K. D., Stergioulas N., 2007, MNRAS, 375, 261

%\bibitem[\protect\citeauthoryear{Sotani et al.}{2017}]{SKTK2017}
%   Sotani H., Kuroda T., Takiwaki T., Kotake K., 2017, Phys.\ Rev.\ D, 96, 063005

\bibitem[\protect\citeauthoryear{Sotani et al.}{2019}]{SKTK2019}
   Sotani H., Kuroda T., Takiwaki T., Kotake K., 2019, Phys.\ Rev.\ D, 99, 123024

\bibitem[\protect\citeauthoryear{Sotani et al.}{2012}]{SNIO2012}
   Sotani H., Nakazato K., Iida K., Oyamatsu K., 2012, Phys.\ Rev.\ Lett., 108, 201101   
   
\bibitem[\protect\citeauthoryear{Sotani et al.}{2013a}]{SNIO2013a}
   Sotani H., Nakazato K., Iida K., Oyamatsu K., 2013a, MNRAS, 428, L21

\bibitem[\protect\citeauthoryear{Sotani et al.}{2013b}]{SNIO2013b}
   Sotani H., Nakazato K., Iida K., Oyamatsu K., 2013b, MNRAS, 434, 2060

\bibitem[\protect\citeauthoryear{Sotani \& Sumiyoshi}{2019}]{SS2019}
   Sotani H., Sumiyoshi K., 2019, Phys.\ Rev.\ D, 100, 083008
   
%\bibitem[\protect\citeauthoryear{Sotani \& Sumiyoshi}{2021}]{SS2021}
%   Sotani H., Sumiyoshi K., 2021, MNRAS, 507, 2766

%\bibitem[\protect\citeauthoryear{Sotani \& Takiwaki}{2016}]{ST2016}
%   Sotani H., Takiwaki T., 2016, Phys.\ Rev.\ D, 94, 044043

%\bibitem[\protect\citeauthoryear{Sotani \& Takiwaki}{2020a}]{ST2020a}
%   Sotani H., Takiwaki T., 2020a, Phys.\ Rev.\ D, 102, 023028 

%\bibitem[\protect\citeauthoryear{Sotani \& Takiwaki}{2020b}]{ST2020b}
%   Sotani H., Takiwaki T., 2020b, MNRAS, 498, 3503

%\bibitem[\protect\citeauthoryear{Sotani \& Takiwaki}{2020c}]{ST2020c}
%   Sotani H., Takiwaki T., 2020c, Phys.\ Rev.\ D, 102, 063025

\bibitem[\protect\citeauthoryear{Sotani, Takiwaki \& Togashi}{2021}]{STT2021}
   Sotani H., Takiwaki T., Togashi H., 2021, Phys.\ Rev.\ D, 104, 123009

\bibitem[\protect\citeauthoryear{Sotani, Tominaga \& Maeda}{2001}]{STM2001}
   Sotani H., Tominaga K., Maeda K. I., 2001, Phys.\ Rev.\ D, 65, 024010

\bibitem[\protect\citeauthoryear{Sotani et al.}{2011}]{SYMT2011}
   Sotani H., Yasutake N., Maruyama T., Tatsumi T., 2011, Phys. Rev. D, 83, 024014

\bibitem[\protect\citeauthoryear{Steiner \& Watts}{2009}]{SW2009}
   Steiner A. W., Watts A. L., 2009, Phys. Rev. Lett., 103, 181101
   
\bibitem[\protect\citeauthoryear{Strohmayer et al.}{1991}]{SHOII1991}
   Strohmayer T., van Horn H. M., Ogata S., Iyetomi H., Ichimaru S., 1991, ApJ, 375, 679

\bibitem[\protect\citeauthoryear{Strohmayer \& Watts}{2005}]{SW2005}
   Strohmayer T. E., Watts A. L., 2005, ApJ, 632, L111

\bibitem[\protect\citeauthoryear{Strohmayer \& Watts}{2006}]{SW2006}
   Strohmayer T. E., Watts A. L., 2006, ApJ, 653, 593

%\bibitem[\protect\citeauthoryear{Thompson \& Duncan}{1993}]{TD1993}
%   Thompson C., Duncan R. C., 1993, Astrophys. J., 408, 194

%\bibitem[\protect\citeauthoryear{Thompson \& Duncan}{1996}]{TD1996}
%   Thompson C., Duncan R. C., 1996, Astrophys. J., 473, 322

\bibitem[\protect\citeauthoryear{Togashi et al.}{2017}]{Togashi17}  % TGTF
   Togashi H., Nakazato K., Takehara Y., Yamamuro S., Suzuki H., Takano M., 2017, Nucl. Phys. A, 961, 78

\bibitem[\protect\citeauthoryear{Torres-Forn\'{e} et al.}{2019}]{TCPOF19}
   Torres-Forn\'{e} A., Cerd\'{a}-Dur\'{a}n P., Passamonti A., Obergaulinger M., Font J. A., 2019, MNRAS, 482, 3967











\end{thebibliography}
\end{document}